\documentclass[11pt,a4paper]{article}
\usepackage[body={16.5cm,25cm}]{geometry}
\usepackage[dvips]{color}
\usepackage{curves}

\usepackage[hang,small,center]{caption}
\usepackage{amsmath,amssymb}
\usepackage{amsfonts}
\usepackage{mathrsfs}
\usepackage{epic}
\usepackage{eepic}
\usepackage{graphicx}
\usepackage{cite}
\newcommand{\ds}{\displaystyle}
\renewcommand{\author}[1]{\large\rm #1\\ \bigskip}
\newcommand{\address}[1]{{\normalsize\it #1\\}\bigskip}
\renewcommand{\title}[1]{\bigskip\bigskip\Large\bf #1\bigskip\bigskip\\}

\newcommand{\Bigpsi}[3]{\phantom{\Psi}_2 \kern -.05em
\Psi_2\left(\genfrac{}{}{0pt}{}{#1}{#2}\biggl|#3\right)}

\def\bea{\begin{eqnarray}}
\def\eea{\end{eqnarray}}
\newcommand{\be}{\begin{equation}}
\newcommand{\ee}{\end{equation}}
\newcommand{\beq}{\begin{equation}}
\newcommand{\eeq}{\end{equation}}
\newcommand{\bedef}{\stackrel{\textrm{def}}{=}}

\newcommand{\R}{{\mathcal{R}}}
\newcommand{\Ru}{\boldsymbol{ R}}

\newcommand{\bosn}{\textrm{\scriptsize $\boldsymbol{N}$}}

\renewcommand{\l}{{\lambda}}
\renewcommand{\i}{{i_1}}
\renewcommand{\j}{{j_1}}
\newcommand{\ip}{{i'_1}}
\newcommand{\jp}{{j'_1}}

\newcommand{\bos}{\boldsymbol{a}}

\newcommand{\kop}{\boldsymbol{k}}
\newenvironment{proof}[1][Proof]{\begin{trivlist}
\item[\hskip \labelsep {\bfseries #1}]}{\end{trivlist}}

\newcounter{app}
\newcounter{sapp}[app]

\begin{document}

\vglue 2 cm
\begin{center}
\title{An integrable 3D lattice model with positive Boltzmann weights}
\author{Vladimir
  V.~Mangazeev$^{1,2}$ ,Vladimir V.~Bazhanov$^{1,2}$ and \\
Sergey M.~Sergeev$^{1,3}$}
\address{$^1$Department of Theoretical Physics,
         Research School of Physics and Engineering,\\
    Australian National University, Canberra, ACT 0200, Australia.\\\ \\
$^2$Mathematical Sciences Institute,\\
      Australian National University, Canberra, ACT 0200,
      Australia.\\\ \\
$^3$Faculty of Education, Science, Technology and Mathematics,\\
University of Canberra, Bruce ACT 2601, Australia.}

\begin{abstract}
In this paper we construct a three-dimensional (3D) solvable lattice
model with non-negative Boltzmann weights.
The spin variables in the model are
assigned to edges of the 3D cubic lattice and run over an infinite
number of discrete states.
The Boltzmann weights satisfy the tetrahedron equation, which is a
3D generalisation of the Yang-Baxter equation.
The weights depend on a free parameter $0<q<1$ and three
continuous field variables. The layer-to-layer transfer matrices of
the model form a two-parameter commutative family. This is the first example
of a solvable 3D lattice model with non-negative Boltzmann
weights.

\end{abstract}

\end{center}

\newpage

\section{Introduction}

The {\em tetrahedron equation}
\cite{Zamolodchikov:1980rus,Zamolodchikov:1981kf} is a
three-dimensional analog of the Yang-Baxter equation. It implies
the commutativity of layer-to-layer transfer matrices
\cite{Bazhanov:1981zm} for
three-dimensional lattice models of statistical mechanics and
field theory and, thus, generalizes the most
fundamental integrability structure of exactly solvable models in two
dimensions \cite{Bax82}.

Historically, the first solution of the tetrahedron equation was
proposed by Zamolodchikov \cite{Zamolodchikov:1980rus,
  Zamolodchikov:1981kf}.  It was subsequently proven by Baxter
\cite{Baxter:1983qc} and further studied in
\cite{BaxterForrester:1985, Baxter:1986phd, BaxterQuispel:1990,
  Sergeev:1995rt,Korepanov:1993, Hietarinta:1994, BaxterBazhanov:1997,
BoosMangazeev:1999}. A generalisation of this solution to any number %$N$
of discrete spin state spins, $N\ge2$, was found in
\cite{Bazhanov:1992jqa,KMS:1993}.  Next, a different solution to the
tetrahedron equations, with spins having infinitely many discrete
states,
was originally constructed in \cite{Bazhanov:2005as} and then
further generalized in \cite{Bazhanov:2008rd} for the case of
continuous spin variables.
Subsequently, all the above solutions were
again rederived from a common point of view based on rather remarkable
geometric considerations \cite{Bazhanov:2010cong}.
It is worth mentioning also that known 3D integrable models
helped to reveal some hidden structures of quantum groups, in particular,
the ``rank-size'' duality \cite{Bazhanov:1992jqa,Bazhanov:2005as}.

Nevertheless, despite
all these fascinating mathematical connections the topic of 3D
integrability has never really
attracted any notable attention in statistical mechanics, since
all solutions of the tetrahedron equations,  hitherto obtained,
always had negative (and even complex) entries
and therefore could not be directly interpreted as Boltzmann weights of
physical model of statistical mechanics.

In this paper we break this unremarkable tradition
and obtain the first
solution of the tetrahedron equation, which has only real non-negative
weights. The spin variables in the model are assigned to the edges of
a 3D cubic lattice and have an infinite number of discrete states, labelled
by non-negative integers. Therefore, every vertex of the lattice can
occur in an infinite number of configurations, determined by spin arrangements
on the six edges attached to the vertex. Not all these arrangements
are allowed, as there are two constraints on the values of the
edge spins at the
vertex (similar to the arrow conservation law in the 2D ice model
\cite{Lieb67}). Forbidden arrangements are assigned with
vanishing weights, however for all allowed ones the weights are real
and positive. The idea of the very existence of such solution was
previously pronounced by one of us in
\cite{Sergeev:2009cl} on the basis of analytical properties of
the Lagrangian function of associated classical integrable discrete systems.

In Sect.2 we present the new solution of the tetrahedron relation
and prove its positivity. Various properties of this solution are
discussed in Sect.3. The vertex weights depend on a single parameter $0<q<1$
and three ``field variables'', similar to those of the 2D six-vertex
model.
The partition function for periodic boundary
condition is defined in Sect.4. The commuting layer-to-layer transfer
matrices are constructed in Sect.5. The ``rank-size'' duality is
considered in Sect.5.4.

\section{A positive solution to the tetrahedron equation}
\subsection{Operator maps of the $q$-oscillator algebras and the
  functional tetrahedron equation}
Remarkably, the derivation of the new solution of the tetrahedron
equation, which
we present here, only requires rather minor modifications to already
existing results \cite{Bazhanov:2005as}.
Consider the $q$-oscillator algebra
\beq\label{q-osc1}
\mathsf{Osc}_{\,q}:\qquad
\kop\,\bos^{\pm}=q^{\pm 1}\,\bos^{\pm}\,\kop\;,\qquad
q\,\bos^+\bos^- - q^{-1} \,\bos^-\bos^+=q-q^{-1},\qquad
\eeq
generated by the three elements
$\kop$, $\bos^+$ and $\bos^-$
and impose an additional relation
\beq
\kop^2=q\,(1 -\bos^+\bos^-)\equiv q^{-1}\,(1-\bos^-\bos^+)\,,
\label{q-osc2}
\eeq
which is consistent with \eqref{q-osc1}.
We will always assume that $0<q<1$ and that the element $\kop$ is
invertible\footnote{The sign of $\kop$ is fixed by the
  representations \eqref{fock-p}, \eqref{fock-m} below.}.
Below we will need to use several matrices acting
in a tensor product of two two-dimensional vector spaces
${\mathbb C}^2\otimes{\mathbb C}^2$. Any such matrix can be
conveniently represented
as a two by two block matrix with two-dimensional blocks where the
matrix indices related to the second vector space numerate the blocks
while the indices of the first space numerate matrix elements inside
the blocks. With these conventions define an operator-valued matrix,
acting in ${\mathbb C}^2\otimes{\mathbb C}^2$,
\beq\label{L-def}
{\boldsymbol L}(\kop,\bos^\pm)=\left(\begin{array}{cccc}
1 & 0 & 0 & 0 \\
0 & \kop & \bos^+ & 0 \\
0 & \bos^- & -\kop & 0\\
0 & 0 & 0 & 1\end{array}\right)\,,\quad
\eeq
whose elements belong to the algebra \eqref{q-osc1}.

In ref.\cite{Bazhanov:2005as} the problem of solving the
tetrahedron equation was reduced to finding matrix representations
of a certain operator map
of the tensor cube of the algebra \eqref{q-osc1}
to itself,
\beq\label{Rmap1}
{\cal R}_{123}:\qquad \mathsf{Osc}_{\,q}\otimes
\mathsf{Osc}_{\,q}\otimes \mathsf{Osc}_{\,q}
\to \mathsf{Osc}_{\,q}\otimes
\mathsf{Osc}_{\,q}\otimes \mathsf{Osc}_{\,q}\,.
\eeq
Let $\kop_i,\bos^\pm_i$, $i=1,2,3$, \ denote the
generators in the first, second and third factors this product,
respectively, and
\beq\label{Rmap2}
\kop_i'={\mathcal R}_{123}\big(\kop_i\big)\,,\qquad
{\bos'}_i^{\,\pm}={\mathcal R}_{123}\big(\bos_i^\pm\big)\,,\qquad i=1,2,3,
\eeq
denote their images under the map \eqref{Rmap1}.

To construct this map introduce
three operator-valued matrices
${\boldsymbol L}_{\alpha,\beta}(\kop_1,\bos^\pm_1)$,
${\boldsymbol L}_{\alpha,\gamma}(\kop_2,\bos^\pm_2)$ and
${\boldsymbol L}_{\beta,\gamma}(\kop_3,\bos^\pm_3)$ acting in a tensor
product of three two-dimensional vector spaces ${\mathbb C}^2\otimes
{\mathbb C}^2\otimes   {\mathbb C}^2$, labelled by $\alpha$, $\beta$
and $\gamma$, respectively. The matrix ${\boldsymbol
  L}_{\alpha,\beta}(\kop_1,\bos^\pm_1)$
 acts non-trivially only in the first two spaces $\alpha$ and $\beta$,
where it is defined  by \eqref{L-def} and coincides with the identity
operator in the third vector space $\gamma$.
Its matrix elements, belong to
the first $q$-oscillator algebra in the direct product \eqref{Rmap1}.
The matrices ${\boldsymbol L}_{\alpha,\gamma}(\kop_2,\bos^\pm_2)$ and
${\boldsymbol L}_{\beta,\gamma}(\kop_3,\bos^\pm_3)$ are defined in a
similar way.

The map \eqref{Rmap1},\eqref{Rmap2} is uniquely defined (up to a sign
of $\kop'_2$) by the
following matrix equation,\footnote{%
This equation is sometimes called the {\em local} Yang-Baxter
equation \cite{Maillet:1989gg}.
Note, that it is not equivalent to the Yang-Baxter equation.
Even thought it has the same matrix structure, the ${\boldsymbol
  L}$-matrices in the LHS and RHS of \eqref{local} are
different. Moreover, they
have operator-valued (rather than
number-valued) matrix elements.}
\beq\label{local}
{\boldsymbol L}_{\alpha,\beta}(\kop_1,\bos^\pm_1)\
{\boldsymbol L}_{\alpha,\gamma}(\kop_2,\bos^\pm_2)\
{\boldsymbol L}_{\beta,\gamma}(\kop_3,\bos^\pm_3)=
{\boldsymbol L}_{\beta,\gamma}(\kop'_3,{\bos'}^\pm_3)\
{\boldsymbol L}_{\alpha,\gamma}(\kop'_2,{\bos'}^\pm_2)\
{\boldsymbol L}_{\alpha,\beta}(\kop'_1,{\bos'}^\pm_1)\,.
\eeq
Explicitly, the map ${\mathcal R}_{123}$ reads
\begin{subequations}
\label{mapping}
\begin{equation}\label{map-eq1}
\begin{array}{rcl}
\ds \kop'_2\,{\bos'}^{\,\pm}_1&=&
\kop^{}_3\bos^\pm_1 \,+\,
\kop^{}_1\bos^\pm_2\bos^\mp_3\,,\\[3mm]
{\bos'}^{\,\pm}_2 &=&
\bos^\pm_1\bos^\pm_3\,-\,
\kop^{}_1\kop^{}_3\bos^\pm_2\,,\\[3mm]
\kop_2'\,{\bos'}^{\,\pm}_3 & =&\ds
\kop^{}_1\bos^\pm_3 \,+\,
\kop^{}_3\bos^\mp_1\bos^\pm_2\,,
\end{array}
\end{equation}
where
\beq\label{map-eq2}
\big(\kop_2'\big)^2
=
\kop_1^2\kop_2^2\kop_3^2
+
\kop_1^{}\kop_3^{}
\big(
q^{-1}\bos_1^+\bos_2^-\bos_3^+ +q\bos_1^-\bos_2^+\bos_3^-
\big)
 +
 \kop_1^2+\kop_3^2
 -
 (q^{-1}+q)\kop_1^2\kop_3^3
\eeq
while ${\kop'}_1$ and ${\kop'}_2$ are given by the relations
\beq\label{map-eq3}
\ds \kop'_1\,\kop'_2=\kop_1\kop_2\,
\qquad\kop'_2\kop'_3=\kop^{}_2\kop^{}_3\,.
\eeq
\end{subequations}

Consider now the direct product of six $q$-oscillator algebras,
\beq\label{Asix}
{\mathcal A}=\mathsf{Osc}_{\,q}\otimes
\mathsf{Osc}_{\,q}\otimes \cdots \otimes \mathsf{Osc}_{\,q}\,.
\eeq
 labelled consequently by
$i=1,2,\ldots,6$ and introduce an abbreviated notation
\beq
{\boldsymbol L}_{\alpha,\beta}^{(i)}=
{\boldsymbol L}_{\alpha,\beta}(\kop_i,\bos^\pm_i)\,\qquad i=1,2,\ldots,6\,.
\eeq
Then, using \eqref{Rmap2} one can rewrite \eqref{local} in the form
\beq\label{local2}
{\boldsymbol L}_{\alpha,\beta}^{(1)}\
{\boldsymbol L}_{\alpha,\gamma}^{(2)}\
{\boldsymbol L}_{\beta,\gamma}^{(3)}=
{\mathcal R}_{123}\Big({\boldsymbol L}_{\beta,\gamma}^{(3)}\
{\boldsymbol L}_{\alpha,\gamma}^{(2)}\,
{\boldsymbol L}_{\alpha,\beta}^{(1)}\Big)\,.
\eeq
which shows that the application of the map ${\mathcal R}_{123}$ is
equivalent to reversing the product of the three ${\boldsymbol
  L}$-operators. It is not difficult to see, that using
\eqref{local2} four times, one can reverse the
order of the following six-fold product
\beq\label{Lsix}
{\boldsymbol L}_{\alpha,\beta}^{(1)}\
{\boldsymbol L}_{\alpha,\gamma}^{(2)}\
{\boldsymbol L}_{\beta,\gamma}^{(3)}\
{\boldsymbol L}_{\alpha,\delta}^{(4)}\
{\boldsymbol L}_{\beta,\delta}^{(5)}\
{\boldsymbol L}_{\gamma,\delta}^{(6)}
=
\mathcal{T}\Big(
{\boldsymbol L}_{\gamma,\delta}^{(6)}\
{\boldsymbol L}_{\beta,\delta}^{(5)} \
{\boldsymbol L}_{\alpha,\delta}^{(4)} \
{\boldsymbol L}_{\beta,\gamma}^{(3)} \
{\boldsymbol L}_{\alpha,\gamma}^{(2)} \
{\boldsymbol L}_{\alpha,\beta}^{(1)} \Big)
\eeq
where the matrices ${\boldsymbol L}_{\alpha,\beta}^{(1)}$,
${\boldsymbol L}_{\alpha,\gamma}^{(2)}$, etc., act in the tensor
product of four vector spaces ${\mathbb C}^2$, labelled $\alpha$,
$\beta$, $\gamma$ and $\delta$, while their matrix elements belong to
the $q$-oscillator algebras \eqref{Asix}. Remarkably, the required
map ${\mathcal T}$ can be decomposed into elementary moves
\eqref{local2} in
two different ways,
\beq\label{Tvar1}
{\mathcal T}= \R_{123}\,\circ\,\R_{145}\,\circ\,\R_{246}\,\circ\,\R_{356}
\eeq
and
\beq\label{Tvar2}
{\mathcal T}=
\R_{356}\,\circ\,\R_{246}\,\circ\,\R_{145}\,\circ\,\R_{123}\,,
\eeq
 Taking into account that the
matrix elements of the product in the LHS of \eqref{Lsix} span
the full basis in \eqref{Asix} one obtains the functional tetrahedron equation
\beq\label{TE-func}
\R_{123}\,\circ\,\R_{145}\,\circ\,\R_{246}\,\circ\,\R_{356}=
\R_{356}\,\circ\,\R_{246}\,\circ\,\R_{145}\,\circ\,\R_{123}\,.
\eeq
Note, that this equation can be verified by direct calculations of
compositions of the maps in both sides, using the explicit expressions
\eqref{mapping}.
For further details of the derivation we refer the reader to the
original publication \cite{Bazhanov:2005as}.

\subsection{Recurrence relations and positivity}
Consider an infinite-dimensional oscillator Fock space,
 spanned by the set of vectors
$|n\rangle$, $n=0,1,2,\ldots,\infty$, \ with the natural scalar product
\beq\label{N-def}
\langle m| n\rangle= \delta_{m,n}\,,\qquad {\boldsymbol N}\,
|n\rangle= n\, |n\rangle\,,\qquad
\langle n| \, {\boldsymbol N} =\langle n| \, n\,,
\eeq
where we have introduced the ``occupation number'' operator $
{\boldsymbol N}$. The algebra \eqref{q-osc1} has two irreducible
highest weight representations acting the this space, which we denote
${\cal F}_q^\pm$.
The representation $\mathcal{F}_q^+$ is defined as
\begin{subequations}\label{fock-p}
\beq
\kop=q^{\bosn+1/2}\,,
\eeq
and
\beq
\begin{array}{lll}
\bos^-|0\rangle=0,\quad &\bos^+|n\rangle=(1-q^{2+2n})|n+1\rangle,\qquad
&\bos^-|n\rangle=|n-1\rangle,
\\[.4cm]
\langle 0|\bos^+=0\,,\qquad& \langle n|\bos^+=\langle n-1|(1-q^{2n}),
&\langle n | \bos^-=\langle n+1|\,,
\end{array}
\eeq
\end{subequations}
with $n=0,1,2\ldots$.
Similarly, the representation $\mathcal{F}_q^-$ is defined as
\begin{subequations}\label{fock-m}
\beq
\kop=q^{-\bosn-1/2},
\eeq
and
\beq
\begin{array}{lll}
\bos^+|0\rangle=0,\qquad
&\bos^-|n\rangle=|n+1\rangle,
&\bos^+|n\rangle=(1-q^{-2n})|n-1\rangle,\\[.4cm]
\langle 0|\bos^-=0\,,
&\langle n | \bos^-=\langle n-1|\,,\qquad
&\langle n|\bos^+=\langle
n+1|(1-q^{-2-2n})\,,
\end{array}
\label{q-osc5}
\eeq
\end{subequations}
with $n=0,1,2\ldots$.
Following \cite{Bazhanov:2005as}
we realise the map \eqref{Rmap2} as an internal
automorphism
\beq\label{internal}
\mathcal{R}_{123}({\boldsymbol x})={\boldsymbol R}_{123}\,{\boldsymbol x}\,
        {\boldsymbol  R}_{123}^{-1}\,,
\qquad {\boldsymbol R}_{123}, {\boldsymbol x}\in \mathsf{Osc}_{\,q}\otimes
\mathsf{Osc}_{\,q}\otimes \mathsf{Osc}_{\,q}\,.
\eeq
of the direct product of the three oscillator algebras. Obviously, there
are eight possible ways, ${\mathcal F}^{\sigma_1}_q\otimes
{\mathcal F}^{\sigma_2}_q\otimes{\mathcal F}^{\sigma_3}_q$,  \ with
$\sigma_1,\sigma_2,\sigma_3=\pm$, \ to  choose a Fock representation
 in this product. Once
the representation is chosen
the matrix elements of the operator $\boldsymbol R$ can be calculated
using the explicit form of
the map \eqref{mapping}. Note, that this procedure uniquely define the matrix
elements of $\boldsymbol R$ (to within
an overall normalization), since the
representations $\mathcal{F}_q^\pm$  are irreducible.
The problem of finding ${\boldsymbol R}$ for the case when all three
representations coincide with
$\mathcal{F}_q^+$ has already been solved in
\cite{Bazhanov:2005as,Bazhanov:2008rd}.
In this paper we consider another symmetric case, when all three
representations coincide with
$\mathcal{F}_q^-$ and demonstrate
some rather remarkable positivity
properties of the resulting operator $\boldsymbol{R}$.

First, using \eqref{mapping}, let us derive recurrence relations
for the matrix elements
\beq\label{mat-def}
R_{\,n_1,\,n_2,\,n_3}^{{\,n'_1,\,n'_2,\,n'_3}^{\phantom{|}}}=
\langle n_1,n_2,n_3\,|\,\boldsymbol{R}\,|\,n'_1,n'_2,n'_3\rangle,\qquad
n_i,n'_i=0,1,2,\ldots\infty,\quad i=1,2,3.
\eeq
of the operator $\boldsymbol{R}$,
where $|n_1,n_2,n_3\rangle=|n_1\rangle\otimes|n_2\rangle\otimes
|n_3\rangle $ denotes states in $\mathcal{F}_q^-\otimes \mathcal{F}_q^-\otimes
\mathcal{F}_q^-$. Eqs.\eqref{map-eq3} imply that the indices $n_i$ and
$n_i'$ obey two constrains
\beq\label{con-law}
n_1+n_2=n_1'+n_2',\qquad n_2+n_3=n_2'+n_3',
\eeq
for all non-zero matrix elements in \eqref{mat-def}. Therefore all
these elements only depend on four independent discrete variables,
for which we choose $n_2$, $q^{-2n_1'}$, $q^{-2n_2'}$ and
$q^{-2n_3'}$. It follows then, that the matrix \eqref{mat-def} can be
represented in the form
\beq
R_{\,n_1,\,n_2,\,n_3}^{{\,n'_1,\,n'_2,\,n'_3}^{\phantom{|}}}=
\delta_{n_1+n_2,n_1'+n_2'}\delta_{n_2+n_3,n_2'+n_3'}
\frac{q^{n_2(n_2+1)-(n_2-n_1')(n_2-n_3')}}
{(q^2;q^2)_{n_2}}
Q_{n_2}(q^{-2n_1'},q^{-2n_2'},q^{-2n_3'}),
\label{rp1}
\eeq
where $\quad n_i, n_i'=0,1,2,3,\ldots$ and we have introduced a set
of (yet unknown) functions $Q_n(x,y,z)$ depending on the three variables
$x=q^{-2n_1'}$, $y=q^{-2n_2'}$ and
$z=q^{-2n_3'}$. The specific $q$-dependent factor in \eqref{rp1}, involving
the Pochhammer symbol
\beq
(x\,;\,p)_n=\prod_{k=0}^{n-1} (1-x\, p^{k})\,,
\eeq
has
been chosen to ensure that the functions $Q_n(x,y,z)$ are polynomials
in $x,y,z$ with coefficients which are themselves polynomials in the
variable $q$ (this immediately follows from \eqref{recur1}, see
below). Next, substituting the formula \eqref{rp1}
into \eqref{internal}, \eqref{Rmap2} and \eqref{mapping},
specialized for the representation \eqref{fock-m},
one can derive a set of recurrence relation for $Q_n(x,y,z)$.
First, consider the simplest case $n_2=0$. Taking the ``$-$'' signs
(lower signs) in
\eqref{map-eq1} and calculating matrix
elements of both sides of these equations sandwiched between the
states $\langle n_1,0,n_3|$ and $|n_1',n_2',n_3'\rangle$, one obtains
a set of three simple relations
\beq
Q_0(x q^{-2},y,z)=
Q_0(x,y q^{-2},z)=
Q_0(x ,y,zq^{-2})=
Q_0(x,y,z).
\eeq
Thus, for the normalization
\beq
R_{0,\,0,\,0}^{0,\,0,\,0}=1\,,
\eeq
one can set
\beq\label{recur2}
Q_0(x,y,z)\equiv 1\,,\qquad \forall\ x,y,z=1,q^{-2},q^{-4},q^{-6}\ldots\,.
\eeq
More generally, Eqs.\eqref{Rmap2} implies the following recurrence
relation,
\beq
Q_{n+1}(x,y,z)=(x-1)\,(z-1)\,Q_n(x\,q^2,y,z\,q^2)
+x\,z\,(y-1)\,q^{2n}\,Q_n(x,y\,q^2,z)\,.\label{recur1}
\eeq
In particular, the next two polynomials read
\beq\label{rp23}
\begin{array}{rcl}
Q_1(x,y,z)&=&1-(x+z)+x\,y\,z\,,\\[.3cm]
Q_2(x,y,z)&=&(1-x)\,(1-x\,q^2)\,(1-z)\,(1-z\,q^2)
-x^2\,z^2\,q^{4}\,(1-y^2)-\\[.3cm]
&&-x\,z\,q^2\,(1+q^2)\,(1-y)\,(1-x-z)\,.
\end{array}
\eeq
Actually, it is not too difficult to solve \eqref{recur1} with the
initial condition \eqref{recur2}
and derive an explicit formula valid for all values of $n$,
\beq
Q_n(x,y,z)=(x;q^2)_n\phantom{|}_2\phi_1
(q^{-2n},\frac{q^{2-2n}}{xy},\frac{q^{2-2n}}{x};q^2,yz\,q^{2n})\label{rp4}
\eeq
where $\phantom{|}_2\phi_1$ is the truncated generalized
hypergeometric function, defined as
\beq
\phantom{|}_2\phi_1(p^{-n},b,c\,;\,p,z)
\,\bedef\,\sum_{k=0}^n\frac{(p^{-n};p)_k\,(b;p)_k}{(p;p)_k\,(c;p)_k}\,z^k\,,
\qquad n\ge0\,.\label{rp6}
\eeq
Let us now formulate our main statement.

\bigskip
\noindent{\bf Theorem.}  {\em For any non-negative integers $n,n_1',n_2',n_3'
\ge0$, and any real $q$ in the interval $0<q<1$, the special values of the polynomials $Q_n$,
\beq
Q_n(q^{-2n_1'},q^{-2n_2'},q^{-2n_3'})\geq0,\qquad \forall\
n,n_1',n_2',n_3'\in {\mathbb Z}_{\ge0}
\eeq
are always non-negative.}
\begin{proof}
First, notice that
for $x,y,z\in\{1,q^{-2},q^{-4},q^{-6},\ldots\}$ the coefficients in front of $Q_n(xq^2,y,zq^2)$
and $Q_n(x,yq^2,z)$ in \eqref{recur1} are non-negative.
Then, a proof by induction simply follows from (\ref{recur1}) and
the initial condition (\ref{recur2}).
\end{proof}

Taking this result into account, one immediately
concludes that all matrix elements of the $R$-matrix given \eqref{rp1}
and \eqref{rp4} are non-negative provided $0<q<1$.

\section{Properties of the $R$-matrix}

\subsection{Matrix elements}
Some care should be taken when calculating the $R$-matrix,
defined by (\ref{rp1}) and (\ref{rp4}), for $n_2>
n'_3$. In this case the third argument of the hypergeometric function
is equal to a non-positive power of $q$, where the function
$\phantom{|}_2\phi_1$ will have a pole.
However, this pole is exactly canceled by a zero coming from
the pre-factor in the RHS of
(\ref{rp4}) and the result will always be finite (see, e.g., the first two
polynomials (\ref{rp23}).
In fact, it is easy to rewrite the formula \eqref{rp1}
in a form which does not have any poles
\beq
\begin{array}{rcl}
R_{\,n_1,\,n_2,\,n_3}^{{\,n'_1,\,n'_2,\,n'_3}^{\phantom{|}}}&=&
\delta_{n_1+n_2,n_1'+n_2'}\>\delta_{n_2+n_3,n_2'+n_3'}\,
q^{n_2(n_2+1)-(n_2-n_1')(n_2-n_3')}
\\[.5cm]
&&\ds\qquad\qquad\times
\sum_{r=0}^{n_2}\frac{(q^{-2n_1'};q^2)_{n_2-r}}{(q^2;q^2)_{n_2-r}}
\frac{(q^{2+2n_1};q^2)_r}{(q^2;q^2)_r}q^{-2r(n_3+n_1'+1)}\label{rp7}
\end{array}
\eeq
A few first matrix elements read,
\beq
\begin{array}{c}
R_{000}^{000}=1,\quad
R_{010}^{010}=q^{-1},\quad R_{110}^{110}=q^{-2},\quad
R_{010}^{101}=q^{-2}-1,\quad R_{121}^{121}=q^{-7}+q^{-5}-q^{-1}\,,
\\[.4cm]
R_{021}^{203}=(q^{-6}-1)(q^{-4}-1),\quad
R_{231}^{231}=q^{-14}+(q^{-6}+q^{-4})(q^{-6}-1)\,.
\end{array}
\eeq
\subsection{Symmetry properties}
Introduce the following constant matrices acting in the direct product
of the three Fock spaces,
\beq
\boldsymbol{P}_{13}\,|n_1,n_2,n_3\rangle =|n_3,n_2,n_1\rangle,\qquad
\boldsymbol{S}_{3}\,|n_1,n_2,n_3\rangle=q^{-n_3^2}\,(q^2;q^2)_{n_3}\,
|n_1,n_2,n_3\rangle \,.
\eeq
The 12-element symmetry group of the $R$-matrix \eqref{rp7}
is generated by two transformations
\beq\label{sym1}
{\boldsymbol{P}}_{13}\, {\boldsymbol{R}}_{123}\, {\boldsymbol{P}}_{13}={\boldsymbol{R}}_{123}\,.
\eeq
and
\beq\label{sym2}
{\boldsymbol{P}}_{12}\,\big({\boldsymbol{R}}_{123}\big)^{t_3}\,
{\boldsymbol{P}}_{12}=q^{\boldsymbol{N}_2-\boldsymbol{N}_1} {\boldsymbol{S}}_3
\,{\boldsymbol{R}}_{123}\,{\boldsymbol{S}}_3^{-1}\,,
\eeq
where the superscript $t_3$ denotes the matrix transposition in the
third space.
\subsection{Tetrahedron equation}
The $R$-matrix (\ref{rp7}) satisfies the tetrahedron equation
\beq\label{TE-op}
\Ru_{123}\,\Ru_{145}\,\Ru_{246}\,\Ru_{356}=
\Ru_{356}\,\Ru_{246}\,\Ru_{145}\,\Ru_{123}\,,
\eeq
which is corollary of \eqref{TE-func} and \eqref{internal}. It involves
operators acting in six Fock spaces, where $\Ru_{ijk}$ acts
as non-trivially in the $i$-th, $j$-th and $k$-th spaces, but acts
as the identity in the other three spaces.
In matrix form Eq.\eqref{TE-op} reads
\beq
\begin{array}{l}
{\ds\sum_{{n'_1,\,n'_2,\,n'_3}\atop {{n'_4,\,n'_5,\,n'_6}^{\phantom{|}}}}}
R^{n'_1\, n'_2\, n'_3}_{{n_1\, n_2\, n_3}^{\phantom{|}}}\
R^{n''_1\,n'_4\,n'_5}_{{n'_1\,n_4\,n_5}^{\phantom{|}}}\
R^{n''_2\,n''_4\,n'_6}_{{n'_2\,n'_4\,n_6}^{\phantom{|}}}\
R^{n''_3\,n''_5\,n''_6}_{{n'_3\,n'_5\,n'_6}^{\phantom{|}}}=\\[.8cm]
\qquad\qquad\qquad={\ds\sum_{n'_1,\,n'_2,\,n'_3\atop
{{n'_4,\,n'_5,\,n'_6}^{\phantom{|}}}}}
R^{n'_3\,n'_5\,n'_6}_{{n_3\,n_5\,n_6}^{\phantom{|}}}\
R^{n'_2\,n'_4\,n''_6}_{{n_2\,n_4\,n'_6}^{\phantom{|}}}\
R^{n'_1\,n''_4\,n''_5}_{{n_1\,n'_4\,n'_5}^{\phantom{|}}}\
R^{n''_1\,n''_2\,n''_3}_{{n'_1\,n'_2\,n'_3}^{\phantom{|}}}\,.
\end{array}\label{rp8}
\eeq
This
equation contains summations over the six
``internal'' indices $n_i'$, running over all
non-negative integer values.
However, due to the presence of two $\delta$-functions in (\ref{rp7})
there are only four independent summations in both sides of
(\ref{rp8}).
Moreover, for any fixed values of the ``external''
indices $n_i$
and $n''_i$ all the summation variables in (\ref{rp8}) are bounded
from above and below.
So, there are no convergence problems in (\ref{rp8}), since all sums
there are finite.

As the reader might have noticed, that Eq.\eqref{rp7} defines a {\em
  constant solution} of the tetrahedron equation \eqref{TE-op},
as all four $R$-matrices therein are exactly the same. Below we will
introduce additional continuous parameters into \eqref{rp7}, which
will play the role of the spectral parameters similar to those in
two-dimensional solvable models.

Let $\lambda_i,\mu_i$, \ $i=1,2,\ldots,6$, be positive real numbers.
Using the conservation laws \eqref{con-law}, it is easy to check that
if $\Ru_{ijk}$ satisfies \eqref{TE-op}, then so does the ``dressed''
$R$-matrices
\beq\label{R-new}
\Ru'_{ijk}=
\left(\frac{\mu_k}{\lambda_i}\right)^{\bosn_j} \Ru_{ijk}^{}
\left(\frac{\lambda_j}{\lambda_k}\right)^{\bosn_i}
\left(\frac{\mu_i}{\mu_j}\right)^{\bosn_k}\;,
\eeq
where the indices $(i,j,k)$ take four sets of values appearing in
\eqref{TE-op}. Note that the twelve parameters $\lambda_i,\mu_i$ enter the four
equations \eqref{R-new} only via eight independent ratios, so these
equations define a solution of \eqref{TE-op} containig
eight continuous parameters. Even though that at the first sight
these new degrees of freedom appear to be trivial, they allow to
define a very non-trivial family of commuting layer-to-layer transfer
matrices (see Sect.5).

In addition to \eqref{R-new} the tetrahedron equation
is, obviously, invariant under diagonal similarity transformations
\begin{equation}
\Ru'_{ijk}= c_i^{\bosn_i}\,c_j^{\bosn_j}\,c_k^{\bosn_k}\,\Ru_{ijk}\,
c_i^{-\bosn_i}\,c_j^{-\bosn_j}\,c_k^{-\bosn_k}.
\label{rp10}
\end{equation}
where $c_1,c_2,\ldots,c_6$ are arbitrary positive constants.
However, these transformations,
will not play any role in the following, since they
are nonessential for periodic boundary conditions.

\subsection{Asymptotic behaviour}

In preparation for considerations of the layer-to-layer transfer
matrices with periodic boundary conditions we need to study an
asymptotic behaviour of the matrix elements of the $R$-matrix \eqref{rp7}
for large values of the indices $n_i, n'_i$. Indeed these indices run
over an infinite number of non-negative integers values, so the convergence
of sums involving these matrix elements will need to be investigated.

Consider the recurrence relation \eqref{recur1} in the limit of large
of large positive values of $n$ and $x,y,z$. Simple estimates
shows that the second term in the RHS \eqref{recur1} will be dominant
in this limit. Then in the leading order one gets,
\beq
Q_{n+1}(x,y,z)\simeq x\,y\,z\,q^{2n}\,Q_n(x,y\,q^2,z)\,.\label{recur3}
\eeq
Solving this equation, estimating corrections from the sub-leading
terms in \eqref{recur1} and substituting the result into \eqref{rp1},
one obtains,
\beq
R_{n_1,n_2,n_3}^{n'_1,n'_2,n'_3}=
\delta_{n_1+n_2,n'_1+n'_2}\>
\delta_{n_2+n_3,n'_2+n'_3}\>q^{-n_1 n_2 -n'_1 n_3' -n_2 n_3+O(n_i)}
\,,\label{ass3}
\eeq
where $n_1,n_2,n_3,n_1',n_2',n_3'\to \infty$ and we have
assumed that these
variables are of the same order of magnitude.
Note, that the leading term in the asymptotics \eqref{ass3}
exactly coincides with the most singular term of the sum \eqref{rp7}
in the limit $q\to0$ (it comes from $r=n_2$ term of the sum).

\section{Partition function }

\begin{figure}[ht]
\setlength{\unitlength}{0.05cm}
\begin{picture}(200,130)(20,-10)
\put(60,50){$R_{i_{l,m,n},\phantom{iiii}j_{l,m,n},\phantom{iiii}k_{l,m,n}}^{i_{l+1,m,n},\,j_{l,m+1,n},\,k_{l,m,n+1}}$}
\put(150,48){\bf\huge$=$}
{\thicklines
\put(180,50){\vector(1,0){80}}
\put(220,10){\vector(0,1){80}}
\put(196,26){\vector(1,1){50}}\put(196.2,26.2){\vector(1,1){50}}}
\put(220,50){\circle*{5}}
\put(188,20){$k_{l,m,n}$}\put(245,80){$k_{l,m,n+1}$}
\put(173,54){$j_{l,m,n}$}\put(258,54){$j_{l,m+1,k}$}
\put(212,3){$i_{l,m,n}$}\put(212,95){$i_{l+1,m,n}$}
\end{picture}
\caption{An arrangement of the edge spin states around the vertex with the
  coordinates $(l,m,n)$. The corresponding Boltzmann weight is
  given by an
element of the $R$-matrix}\label{vertex}
\end{figure}
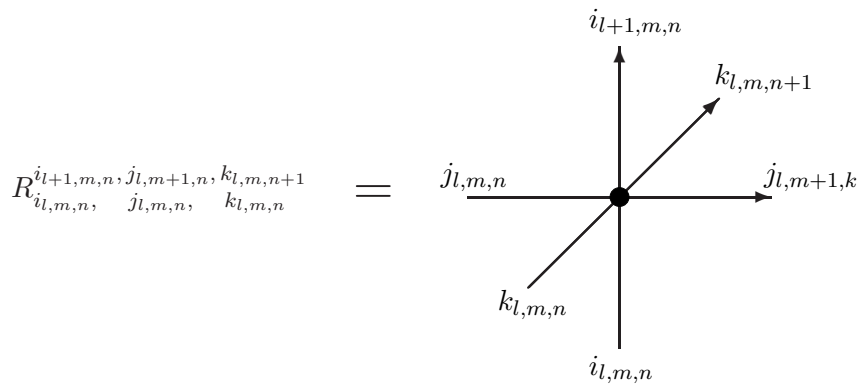

In this Section we define a solvable 3D model of statistical mechanics
with non-negative Boltzmann weights.
Consider the cubic lattice of a size $L\times M\times N$, with
the height $L$, the width $M$ and the depth $N$. Here we assume
that the lattice axes are oriented along the vertical and
two horizontal directions, ``left to right'' and ``front to back''.
The vertices of the lattice are labelled by the
coordinates $(l,m,n)$, where \ $l=1,...,L$, $m=1,...,M$, $n=1,...,N$.
The edges of the lattice carry fluctuating spin variables
taking arbitrary non-negative integer values.
In the previous Sections these spin
variables (oscillator occupation numbers in the Fock spaces) were
denoted $n_1,n_2,n_3,\ldots$. Here it will be more convenient to
use the symbols
$i,j,k$, indexed by the coordinates of the adjacent vertex, as shown
in Fig.\ref{vertex}. The spins $i_{l,m,n}$ are associated with
the vertical edges ($i$-type spins), the spins $j_{l,m,n}$
with the horizontal left-to-right edges ($j$-type spins)
and the spins $k_{l,m,n}$ with
the horizontal front-to-back edges ($k$-type spins).
Each vertex configuration is assigned with a
Boltzmann weight given by an element of the $R$-matrix \eqref{rp7},
also shown in Fig.\ref{vertex}. Actually, we will use the
``dressed'' solution \eqref{R-new} of the tetrahedron equation, which
includes additional edge weights. However, due to multiple
conservation laws for the spin variables (see below) these edge weights
can be equivalently redistributed
among different edges of the lattice and it is not
necessary to have them for all the edges.
For our purposes it will be
convenient to keep the edge weights only on the boundary
edges\footnote{%
The situation is similar to the 2D six-vertex model, where one can
move vertical and horizontal fields to a single column and a single row
of the lattice.}.

In this paper we consider the case of the periodic boundary
conditions in all three directions
\beq
i_{1,m,n}=i_{L+1,m,n}\;,\qquad j_{l,1,n}=j_{l,M+1,n}\;,
\qquad k_{l,m,1}=k_{l,m,N+1}\,.\label{PBC}
\eeq
First, note that the $\delta$-functions in (\ref{rp7}) lead to the following
local conservation laws for each vertex
\beq
i_{l,m,n}+j_{l,m,n}=i_{l+1,m,n}+j_{l,m+1,n},\qquad j_{l,m,n}+k_{l,m,n}=j_{l,m+1,n}+k_{l,m,n+1},\quad \forall\>
l,m,n\,. \label{ass4}
\eeq
in any allowed spin arrangement on the whole lattice.
As an immediate consequence there will be many ``global''
conservation laws for various sums of spins on 1D chains and 2D layers
of edges of the same type. For example, the set of all horizontal coordinate
planes divides the whole lattice into $L$ layers. Each of these
layers will contain $M\times N$ vertical edges with the spins
$\{i_{l,m,n}\}$, $m=1,\ldots,M,\,n=1,\ldots,N$. Then for any allowed
spin arrangement on the whole lattice
the sum of these $i$-type spins
\beq
\mathcal{I}=|\{i\}|=
\sum_{n=1}^N \sum_{m=1}^M i_{l,m,n}\;,\label{I-def}
\eeq
will be the same for all horizontal layers, i.e., it will not depend
on the vertical coordinate $l$. Similarly, define another two sums
\beq
\mathcal{J}=\sum_{n=1}^N \sum_{l=1}^L j_{l,m,n}\;,\qquad
\mathcal{K}=\sum_{m=1}^M \sum_{l=1}^L k_{l,m,n}\;.
\eeq
for the $j$-type and $k$-type spins. In addition,
we will also use the following 1D sums of spins
\beq
\begin{array}{rclrclrcl}
\mathcal{I}^{(M)}_{n}&=&\ds\sum_{m=1}^{M}i_{l,m,n},\quad
&\mathcal{J}^{(L)}_{n}&=&\ds\sum_{l=1}^{L} j_{l,m,n},\quad
&\mathcal{K}^{(L)}_{m}&=&\ds\sum_{l=1}^{L}k_{l,m,n},\\[.4cm]
\mathcal{I}^{(N)}_{m}&=&\ds\sum_{n=1}^{N}i_{l,m,n},
&\mathcal{J}^{(N)}_{l}&=&\ds\sum_{n=1}^{N}l_{l,m,n},
&\quad \mathcal{K}^{(M)}_{l}&=&\ds\sum_{m=1}^{M}k_{l,m,n}
\end{array}\label{sum-def}
\eeq
Note that due \eqref{ass4} the above sums depend only on one
coordinate, instead of two. For instance, the first of these sum
$\mathcal{I}^{(M)}_n$ does not depend on $l$. This means that the sum of spins
on a row of vertical edges, obtained from each other by translations
in the front-to-back direction,
does not depend on the height of this row in
the lattice. Equipped with these definitions, introduce a function of spins
\beq
U=\sum_{l=1}^{L} \mathcal{J}^{(N)}_l\,\mathcal{K}^{(M)}_l
+ \sum_{m=1}^M \mathcal{I}^{(N)}_m\,
\mathcal{K}^{(L)}_m
+ \sum_{n=1}^N \mathcal{I}^{(M)}_n\, \mathcal{J}^{(L)}_n\,,\label{U-def}
\eeq
which is expressed only in terms of the above spin sums.

Remind, that the spins in the model run over an infinite number of
values (all non-negative integers),
therefore, there could be potential convergence problems for the
partition function with the  periodic
boundary conditions \eqref{PBC}.
To better understand the situation let us estimate the leading
asymptotics of the product of the vertex weights over all lattice
vertices,
\beq
{\mathcal P}=\prod_{l,m,n}
R_{i_{l,m,n},\phantom{iiii}j_{l,m,n},
\phantom{iiii}k_{l,m,n}}^{i_{l+1,m,n},\,j_{l,m+1,n},\,k_{l,m,n+1}}\,,
\label{P-def}
\eeq
for a generic spin configuration, when all the spins are large
\beq
i_{i,m,n}\sim j_{l.m,n}\sim k_{l,m,n}\sim O(\Lambda),\qquad \Lambda\to \infty\,,
\label{large}
\eeq
but kept of the
same order of magnitude,
so that their ratios remain finite.
Using the asymptotics \eqref{ass3}, the periodic boundary condition
\eqref{PBC} and the local conservation laws \eqref{ass4},
one can show that
\beq
{\log {\mathcal P}}/ {\log q}\,\sim\,-U + S\big(\{d\}\big)
+O(\Lambda)\,,\label{P-asym}
\eeq
where $U$ is defined in \eqref{U-def} and the second term
\beq
S\big(\{d\}\big)=\sum_{l,m,n}\Big\{\sum_{s=1}^{m-1} d_{l,s,n} \Big(
 \sum_{r=1}^{l-1} d_{r,m,n} +\sum_{t=1}^{n-1} d_{l,m,t}\Big)-
 \sum_{r=1}^{l} d_{r,m,n} \ \sum_{t=1}^{n} d_{l,m,t}\Big\}\,,\label{S-def}
\eeq
depends only on a set of differences of the spins
\beq
d_{l,m,n}\;=\;i_{l+1,m,n}-i_{l,m,n}\;=\;j_{l,m,n}-j_{l,m+1,n}
\;=\;k_{l,m,n+1}-k_{l,m,n}\,.\label{d-def}
\eeq
Remarkably, thanks to \eqref{ass4}, there are three alternative expressions
for the above differences, so that they can be solely associated with either
$i$-type, $j$-type or $k$-type spins.
 Also, it is worth noting that
the quantity \eqref{S-def} can be written in the form
\beq
S\big(\{d\}\big)=2\sum_{\mathcal C}
\ d_{l_1,m_1,n_1} d_{l_2,m_2,n_2}\,,\label{S-def2}
\eeq
where sum is taken over a set coordinates satisfying the conditions
\beq\label{C-def}
{\mathcal C}: \qquad
1\le l_1\le l_2 \le( L-1),\qquad
1\le m_2\le m_1 \le (M-1),\qquad
1\le n_1\le n_2 \le (N-1)\,.
\eeq

Now, we are ready to define the partition function of the model.
First, define a restricted partition function,
 \beq
 \mathcal{Z}_\mathcal{I}=\sum_{|\{i\}|=\mathcal{I}}\ \sum_{\{j,k\}}
 q^{\/\mu U}\,v^\mathcal{J} \,w^\mathcal{K} \prod_{l,m,n}
\Big(q^{j_{l,m,n}}\ R_{i_{l,m,n},\phantom{iiii}j_{l,m,n},\phantom{iiii}k_{l,m,n}}^{i_{l+1,m,n},
\,j_{l,m+1,n},\,k_{l,m,n+1}}\Big)\,, \qquad I=0,1,2,\ldots\label{ZI-def}
\eeq
where the sum is taken over all states of $j$- and $k$-type spins
but the sum over $i$-spins is restricted to configurations satisfying
the condition $|\{i\}|=\mathcal{I}$, which fixes
the total sum of $i$-spins in any horizontal layer (cf. \eqref{I-def}).
Here $v,w$ are arbitrary (positive) parameters, simply related to the edge
weights in \eqref{R-new}. Actually, we found it convenient to also include
some of these edge factors in the last product in \eqref{ZI-def} to fulfill
some spatial requirements, required for Eqs.\eqref{t1}, \eqref{t2}
below.
Moreover, we have included an additional term $q^{\mu U}$, containing a new
(real) parameter $\mu$. The purpose of this term is to regularize the sum
for large values of spin variables. Indeed, with an account of
\eqref{P-asym}, the large-spins asymptotics \eqref{large} of the summand in
\eqref{ZI-def} reads
\beq
\mbox{summand in \eqref{ZI-def}}= q^{(\mu-1) U +
  S(\{d\})+O(\Lambda)}\,.
\label{summ-ass}
\eeq
Note, that the quantity $U$ is positive, it is quadratic in spins and
diverge like $U\sim O(\Lambda^2)$. The second term $S(\{d\})$, is also
quadratic in spins, but for a fixed value of $\mathcal{I}$ it remains finite,
when $\Lambda\to\infty$. Indeed, according to \eqref{S-def} and
\eqref{d-def} this term can be expressed only in terms of differences
of the $i$-type spin. However, since the total sum of these spins is
fixed to $\mathcal{I}$, one concludes that $S(\{d\})\sim
O(\mathcal{I}^2)$,
independent of
$\Lambda$. Thus, if the parameter $\mu>1$ the summand in
the formula \eqref{ZI-def} vanishes exponentially for large spins (remind, that
$q<1$) and the sum over $j$- and $k$-type spins therein will
converge. Next, for $\mu=1$ there might be growing terms in the
exponent of \eqref{summ-ass}, which are linear in spin. However, such
terms are not dangerous, since they can be dumped by
choosing sufficiently small parameters $v$ and $w$ in
\eqref{ZI-def}. More detailed estimates suggest that \eqref{ZI-def}
converges for
\beq
\mu=1\;,\quad v<1\;,\quad w<1\;.
\eeq

In the next sections we will show that the partition function \eqref{ZI-def}
corresponds to an integrable 3D model, in the sense that the corresponding
layer-to-layer transfer matrices form a two-parameter commutative family.
The full partition function is defined
\beq
{\mathcal Z}=\sum_{{\mathcal I}=0}^\infty \, u^\mathcal{I}\, {\mathcal
  Z}_{\mathcal I}
\,,\qquad u<1\label{Z-def}
\eeq
where the parameter $u$ is related to the vertical edge weights.
The convergence of the sum \eqref{Z-def}
requires an additional study (it could require an additional
$I$-dependent dumping factor).

\section{Commuting family of layer-to-layer transfer matrices}
\subsection{Definition of the transfer matrix}
\begin{figure}
\setlength{\unitlength}{0.03cm}
\begin{picture}(500,200)(-20,-30)
\multiput(0,0)(60,0){3}{
{\thicklines
\put(80,25){\vector(1,0){60}}\put(120,65){\vector(1,0){60}}
\put(105,0){\vector(0,1){50}}\put(145,40){\vector(0,1){60}}
\put(194,89){\vector(0,1){50}}\put(169,114){\vector(1,0){60}}
\put(89,9){\vector(1,1){70}}}\put(178,98){\vector(1,1){35}}
\multiput(163,83)(5,5){3}{\circle*{1.5}}
}
\multiput(270,25)(8,0){3}{\circle*{1.5}}
\multiput(310,65)(8,0){3}{\circle*{1.5}}
\multiput(359,114)(8,0){3}{\circle*{1.5}}
\put(215,0){
{\thicklines
\put(80,25){\vector(1,0){60}}\put(120,65){\vector(1,0){60}}
\put(105,0){\vector(0,1){50}}\put(145,40){\vector(0,1){60}}
\put(194,89){\vector(0,1){50}}\put(169,114){\vector(1,0){60}}
\put(89,9){\vector(1,1){70}}}\put(178,98){\vector(1,1){35}}
\multiput(163,83)(5,5){3}{\circle*{1.5}}
}
\put(95,-10){\footnotesize$i_{1,1}$}
\put(155,-10){\footnotesize$i_{2,1}$}
\put(215,-10){\footnotesize$i_{3,1}$}
\put(305,-10){\footnotesize$i_{M,1}$}
\put(55,22){\footnotesize$j_{1,1}$}
\put(97,62){\footnotesize$j_{1,2}$}
\put(140,115){\footnotesize$j_{1,N}$}
\put(210,135){\footnotesize$k_{1,1}$}
\put(270,135){\footnotesize$k_{2,1}$}
\put(330,135){\footnotesize$k_{3,1}$}
\put(425,135){\footnotesize$k_{M,1}$}
\end{picture}
\caption{The layer-to-layer transfer matrix}
\label{fig-tm}
\end{figure}
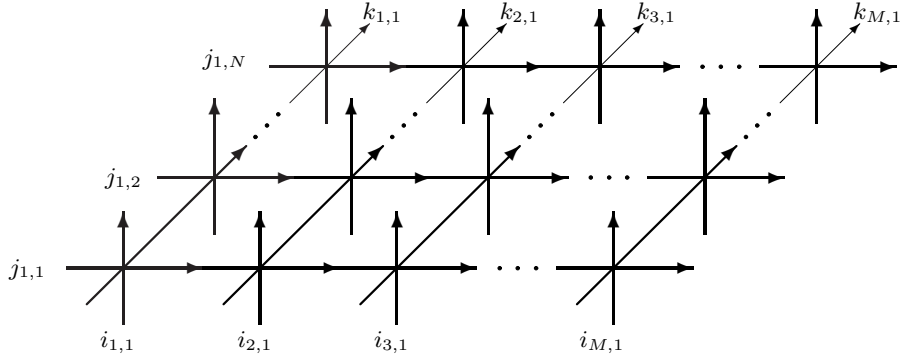

The purpose of this section is to define a commuting family
of layer-to-layer transfer
matrices, associated with the partition function
\eqref{ZI-def}. Consider a particular horizontal layer of the lattice
shown in Fig.\ref{fig-tm}, corresponding to some fixed value of the
height $l$ and assume periodic boundary conditions in both horizontal
directions.
Redenote the spins, associated with this layer, by
dropping the coordinate $l$ from the indices,
\beq
i_{l,m,n}\to i_{m,n},\quad
i_{l+1,m,n}\to i'_{m,n},\quad
j_{l,m,n}\to j_{m,n},\quad
k_{l,m,n}\to k_{m,n}
\eeq
The layer-to-layer transfer matrix is defined as
\beq
T_{\{i\>\}}^{\{i'\}}(v,w)=\sum_{\{j,k\}}
 q^{\/\mu J K} \,v^J \,w^K
\Big(\prod_{m} q^{\mu\,\mathcal{I}^{(N)}_m k_{m,1}}\Big)
\Big(\prod_{n} q^{\mu\,\mathcal{I}^{(M)}_n j_{1,n}}\Big)
\prod_{m,n} \Big(q^{j_{m,n}}\
R_{i_{m,n},\phantom{iiii}j_{m,n},\phantom{iiii}k_{m,n}}^{i'_{m,n},
\,j_{m+1,n},\,k_{m,n+1}}\Big)\,, \label{T-def}
\eeq
where $\mathcal{I}^{(N)}_m$ and $\mathcal{I}^{(M)}_n$ are defined in
\eqref{sum-def} and
\beq
J=\sum_{n=1}^N j_{m,n},\qquad
K=\sum_{m=1}^M k_{m,n},\label{JK-def}
\eeq
Note that for the periodic boundary conditions in horizontal
directions $J$ is independent of
$m$, $K$ is independent of $n$, while  $\mathcal{I}^{(N)}_m$ and
$\mathcal{I}^{(M)}_n$ are the same for all horizontal layers.
Let $\langle \Psi_{\mathcal{I}} |$  and $| \Psi_{\mathcal{I}} \rangle$
be vectors, describing the superposition (with coefficient one)
of all $i$-type spin states, obeying the total sum constraint
$|\{i\}|=\mathcal{I}$. Then the partition function \eqref{ZI-def} can
be written as
\beq
\mathcal{Z_I}=\langle \Psi_{\mathcal{I}}|\,\boldsymbol{T}(v,w)^L\,|
\Psi_{\mathcal{I}} \rangle
\eeq
Below we will show that the transfer matrices \eqref{T-def} form a
two-parameter commutative family,
\beq\label{tcomm}
[\,\boldsymbol{T}(v,w)\,,\,\boldsymbol{T}(v',w')\,]=0\,,\qquad \forall \ v,w,v',w'
\eeq

\subsection{Composite Yang-Baxter equation}
\begin{figure}[ht]
\begin{picture}(200,180)(-100,-20)
{\thicklines
\put(80,25){\vector(1,0){50}}\put(120,65){\vector(1,0){50}}
\put(105,0){\vector(0,1){50}}\put(145,40){\vector(0,1){50}}
\put(194,89){\vector(0,1){50}}\put(169,114){\vector(1,0){50}}
\put(89,9){\vector(1,1){70}}}\put(178,98){\vector(1,1){35}}
\multiput(163,83)(5,5){3}{\circle*{1.5}}
\put(214,133){\footnotesize$k_{1}^{}$}\put(219,117){\footnotesize$j_N'$}
\put(148,117){\footnotesize$l_{N}^{}$}
\put(196,85){\footnotesize$i_N^{}$}\put(177,140){\footnotesize$i_N'$}
\put(156,100){\footnotesize$k_N^{}$}
\put(161,76){\footnotesize$k_3^{}$}\put(118,34){\footnotesize$k_2^{}$}
\put(80,4){\footnotesize$k_1^{}$}
\put(66,22){\footnotesize$j_1^{}$}\put(130,18){\footnotesize$j_1'$}
\put(107,-2){\footnotesize$i_1$}\put(95,50){\footnotesize$i_1'$}
\put(108,62){\footnotesize$j_2^{}$}\put(170,58){\footnotesize$j_2'$}
\put(147,38){\footnotesize$i_2^{}$}\put(135,90){\footnotesize$i_2'$}
\end{picture}
\caption{A front-to-back line of the cubic lattice}\label{rmat-fig}
\end{figure}
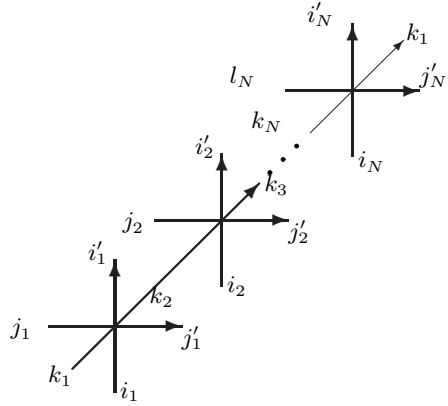

It is well known that any edge-spin model on the cubic lattice can be
viewed as a two-dimensional model on the square lattice with
an enlarged space of states for the edge spins
(see \cite{Bazhanov:2005as} for additional explanations).
Consider a line of vertices in the front-to-back direction and let
\begin{equation}
\boldsymbol{i}=\{i_1^{},i_2^{},\dots,i_N^{}\}\;,\quad
\boldsymbol{i}'=\{i_1',i_2',\dots,i_N'\}\;,\quad\textrm{etc.}
\end{equation}
denote multi-spin variables, describing the states of external edges
of similar types, as shown in Fig.\ref{rmat-fig}. Also, let
$k_1,k_2,\ldots,k_N$ denote states of the internal edges along the
line in the front-to-back direction, where we assume the periodic
boundary conditions $k_{N+1}=k_1$. Also, it is useful to introduce
the variables
\beq
I=\sum_{n=1}^N i_n\;,\quad I'=\sum_{n=1}^N i_n'\;,\quad
J=\sum_{n=1}^N j_n\;,\quad J'=\sum_{n=1}^N j_n'\;.
\eeq
Define a composite weight
\beq
\mathbb{S}_{\boldsymbol{i}\,\,\boldsymbol{j}}^{\boldsymbol{i}'
  \boldsymbol{j}'}(w)=
\sum_{\{k\}} w^{k_1}
\prod_{n=1}^{N}R_{i_n^{},j_n^{},k_n^{}}^{i_n',j_n',k_{n+1}^{}}
\label{compR}
\eeq
where $w$ is an arbitrary (positive) parameter.
The presence of the
delta functions in \eqref{rp7} lead to two ``global'' conservation laws for the
multi-spin variables
\beq\label{IJcons}
I=I',\qquad
J=J'\,,
\eeq
and also determine a local structure of non-zero matrix
elements of $\mathbb{S}$,
\begin{equation}
\mathbb{S}_{\boldsymbol{i}^{},\boldsymbol{j}^{}}^{\boldsymbol{i}',\boldsymbol{j}'}(w)\;\sim\; \mbox{const}\,
\prod_{n=1}^N \delta_{i_n^{}+j_n^{},i_n'+j_n'}\,.\label{6vertex}
\end{equation}
Standard arguments \cite{Bazhanov:1981zm}
relating the tetrahedron and Yang-Baxter
equations allows one to conclude that the composite
$R$-matrix \eqref{compR} satisfies the Yang-Baxter equation
\beq\label{YBE-c}
\sum_{\{\boldsymbol{i}',\boldsymbol{j}',\overline{\boldsymbol{j}}'\}}
\mathbb{S}_{\boldsymbol{i}^{},
  \boldsymbol{j}^{}}^{\boldsymbol{i}',\boldsymbol{j}'}(w)\ \
\mathbb{S}_{\boldsymbol{i}',
\overline{\boldsymbol{j}}^{}}^{\boldsymbol{i}'',\overline{\boldsymbol{j}}'}(w')\ \
\mathbb{S}_{\boldsymbol{j}',
\overline{\boldsymbol{j}}'}^{\boldsymbol{j}'',\overline{\boldsymbol{j}}''}(w'/w)=
\sum_{\{\boldsymbol{i}',\boldsymbol{j}',\overline{\boldsymbol{j}}'\}}
\mathbb{S}_{\boldsymbol{j}^{},
\overline{\boldsymbol{j}}^{}}^{\boldsymbol{j}',\overline{\boldsymbol{j}}'}(w'/w)\ \
\mathbb{S}_{\boldsymbol{i}^{},
\overline{\boldsymbol{j}}'}^{\boldsymbol{i}',\overline{\boldsymbol{j}}''}(w')\ \
\mathbb{S}_{\boldsymbol{i}',
\boldsymbol{j}'}^{\boldsymbol{i}'',\boldsymbol{j}''}(w)\;.
\eeq
This equation states an equality of two linear operators,
acting in a direct product of three
identical infinite-dimensional vector spaces
$({\mathcal F}^-_q)^{\otimes N}\otimes
({\mathcal F}^-_q)^{\otimes N}\otimes ({\mathcal F}^-_q)^{\otimes N}$,
spanned on the vectors
\beq
|i_1,\ldots,i_N\rangle\otimes |j_1,\ldots,j_N\rangle\otimes
|\overline{j}_1,\ldots,\overline{j}_N\rangle\,,
\qquad i_n,j_n,\overline{j}_n=0,1,2,\ldots\infty\,,\qquad n=1,2,\ldots,N\,.
\eeq
The relations \eqref{IJcons} imply a conservation of sums of spins
in each of the three spaces
\beq\label{YBE-split}
I=I'=I'',\qquad J=J'=J'', \qquad \overline{J}=\overline{J}'=\overline{J}''\,.
\eeq
Therefore, Eq.\eqref{YBE-c} reduces to a
direct sum of an infinite number of Yang-Baxter equations,
corresponding to particular values of $I,J,\overline{J}=0,1,2,\ldots,\infty$.
Further, with \eqref{6vertex} it is easy to see that Eq.\eqref{YBE-c}
is not affected by the replacement
\beq\label{adfield}
\mathbb{S}_{\boldsymbol{i}^{},
  \boldsymbol{j}^{}}^{\boldsymbol{i}',\boldsymbol{j}'}(w)\ \
\to
(q^{\mu I} v/v_n)^{j_{1,n}}\
\mathbb{S}_{\boldsymbol{i}^{},
  \boldsymbol{j}^{}}^{\boldsymbol{i}',\boldsymbol{j}'}(w),\qquad
\mathbb{S}_{\boldsymbol{i}^{},
  \overline{\boldsymbol{j}}^{}}^{\boldsymbol{i}',
\overline{\boldsymbol{j}}'}(w')\ \
\to
(q^{\mu I} v/v_n)^{j_{1,n}}\
\mathbb{S}_{\boldsymbol{i}^{},
 \overline{
   \boldsymbol{j}}^{}}^{\boldsymbol{i}',\overline{\boldsymbol{j}}'}(w')\,.
\eeq
usually referred to as an introduction of ``horizontal fields''.

From the results of \cite{Bazhanov:2005as} it is clear that the
composite $R$-matrix
\eqref{compR} should be closely related to the $R$-matrices associated
with the affine quantum algebra $U_q(\widehat{sl}(N))$.
These models were discovered in the early 1980s \cite{Cherednik:1980o,
Schultz:1981s,Babelon:1981s,Perk:1981n} and have since
found numerous applications in integrable systems. They are related to
an anisotropic deformation of the sl(n)-invariant Heisenberg magnets
\cite{Yang:1967s,Uimin:1970o,Lai:1974l,Sutherland:1975m}.
In the simplest $N=2$ case, these models include the most
general six-vertex model \cite{Baxter:1971g}
and all its higher-spin descendants.
Indeed,
following the arguments of \cite{Bazhanov:2005as} one can identify the
subspace
\beq
\pi_I\;=\;\{i_1,i_2,\dots,i_N\;:\;\;\sum_{n=1}^N i_n = I\;\}
\eeq
with the rank $I$ symmetric tensor representation of
$U_q(\widehat{sl}(N))$.
More detailed analysis shows that the composite $R$-matrix
\eqref{compR} can be viewed as an infinite direct sum:
\begin{equation}\label{decomp}
\mathbb{S}(w) \;=\; \mathop{\bigoplus}_{I,J=0}^\infty \ {\mathbf R}_{I,J}^{(sl_N)}(w)
\end{equation}
of the $U_q(\widehat{sl}(N))$
$R$-matrices, ${\mathbf R}_{I,J}^{(sl_N)}$,
intertwining the symmetric tensor
representations
$\pi_I$ and $\pi_J$. It is worth noting that in this setting these
matrices have some specific normalization
uniquely determined by the definition \eqref{compR} and
the solution of the tetrahedron equation \eqref{rp7}.

As an illustration consider the case $N=2$. Then using
\eqref{rp7}, \eqref{compR} and \eqref{decomp} one obtains\footnote{In
  writing \eqref{RIJ-def} we have changed
the overall normalization factor.} \cite{BMS13},
\beq\label{RIJ-def}
\begin{array}{l}
\ds\big[R_{I,J}^{(sl_2)}(\l)
\big]_{\i,\j}^{\ip,\jp}\,=\,\ds\delta_{\i+\j,\ip+\jp}
\frac{q^{i_1^2+(I-\i)(J-\jp)-\ip(\ip-\j)+2I+\frac{1}{2}IJ-\frac{1}{2}m(I,J)}}
{(q^2;q^2)_{\i}\,(q^2;q^2)_{I-\i}}
\\[.6cm]
\ds\qquad\times\  \l^{\i-\ip-m(I,J)}\
(\l^2 q^{-I-J};q^2)_{m(I,J)+1}\
\sum_{k=0}^{\i}\sum_{l=0}^{I-\i}
\frac{(-1)^{k+l}\,q^{2k(\ip-\j)-2l(J-I-\j+\i)}}
     {q^{k(k+1)+l(l+1)}\ (1-\l^2q^{\,I-J-2k-2l})^{\phantom{A^A}} }\\[.8cm]
\ds\qquad\times\ \frac{(q^{-2\i},q^{2+2\j};q^2)_k\
(q^{-2\jp};q^2)_{\i-k}}{(q^2;q^2)_k}
\frac{
(q^{-2(I-\i)},q^{2(1+J-\j)};q^2)_l\ (q^{-2(J-\jp)};q^2)_{I-\i-l}}{(q^2;q^2)_l}
\end{array}
\eeq
where $m(i,j)=\min(i,j)$ and
\beq
w=\lambda^2,\qquad
0\le \i,\ip\le I,\qquad
0\le \j,\jp\le J\,.
\eeq
This is a general expression for the
``higher-spin'' $R$-matrix of the six-vertex model,
with $(I+1)$- and $(J+1)$-state spins on the vertical and horizontal
edges, respectively.
Note, in particular, that for $I=J=1$
the formula \eqref{RIJ-def} reduces to the
$R$-matrix of the six-vertex model \cite{Bax72},
\beq\label{6v}
\begin{array}{rcccc}
\big[R_{1,1}^{(sl_2)}(\lambda)\big]_{00}^{00}&=&
\big[R_{1,1}^{(sl_2)}(\lambda)\big]_{11}^{11}&=&q \lambda-
(q\lambda)^{-1}\,,\\[.3cm]
\big[R_{1,1}^{(sl_2)}(\lambda)\big]_{10}^{10}&=&
\big[R_{1,1}^{(sl_2)}(\lambda)\big]_{01}^{01}&=&\lambda-\lambda^{-1}\,,\\[.3cm]
\big[R_{1,1}^{(sl_2)}(\lambda)\big]_{10}^{01}&=&
\big[R_{1,1}^{(sl_2)}(\lambda)\big]_{01}^{10}&=&q -q^{-1}\,.
\end{array}
\eeq
The derivation of \eqref{RIJ-def} and its connections to other
solutions of the Yang-Baxter equation, related with the six-vertex
model are given in \cite{BMS13}.

\subsection{Inhomogeneous case and commutativity}

Let us again refer to Fig.\ref{fig-tm} and introduce multi-spins variables
\beq
\boldsymbol{i}_m=\{i_{m,1},i_{m,2},\ldots,i_{m,N}\},\qquad
\boldsymbol{j}_m=\{j_{m,1},j_{m,2},\ldots,j_{m,N}\}\,,\quad{\rm etc.}
\eeq
describing states of spins on lines of similar edges, extended in
the front-to-back direction. Introduce also two set of positive real
numbers, $\{v\}=\{v_1,v_2,\ldots,v_n\}$ and $\{w\}=\{w_1,w_2,\ldots,w_m\}$, such that
\beq
v_1v_2\cdots v_n=1,\qquad w_1w_2\cdots w_m=1\,.
\eeq
These numbers will parameterize inhomogeneities of the model.
Consider the transfer matrix,
\beq\label{TM}
T_{\{i^{}\}}^{\{i'\}}(v,w\,|\,\{v\},\{w\})
\;=\;
\sum_{\{\boldsymbol{j}\}}q^{JM}\,
\Big(\prod_n q^{\mu_3 \mathcal{I}^{(M)}_n j_{1,n}} (v/v_n)^{j_{1,n}}\Big)
\prod_{m=1}^M
\mathbb{S}_{\boldsymbol{i}_m^{},\boldsymbol{j}_m^{}}^{\boldsymbol{i}_m',\boldsymbol{j}_{m+1}^{}}(q^{\mu_2\mathcal{I}_m^{(N)}+\mu_1
  J}\ w/w_m)
\eeq
where $\mathcal{I}^{(N)}_m$ and $\mathcal{I}^{(M)}_n$ are defined in
\eqref{sum-def}, $J$ is defined in \eqref{JK-def} and the constants
$\mu_1,\mu_2,\mu_3>1$ are real. It is not difficult to see that
\eqref{TM} reduces to \eqref{T-def} if
\beq
v_1=v_2=\ldots =v_n=1,\qquad
w_1=w_2=\ldots =w_m=1,\qquad \mu_1=\mu_2=\mu_3=\mu.
\eeq
The parameters in \eqref{TM}
have the following interpretation from the point of
view of a 2D lattice model with the composite weights \eqref{compR}.
 The parameter $w$ is the
spectral parameter, associated with the horizontal direction.
The constants $w_m$ provide a set of spectral parameters, associated
with the vertical direction
(usually called inhomogeneities of the spectral
parameter). According to the decomposition \eqref{decomp} the
transfer matrix is infinite sum of various ``fusion'' transfer
matrices with
different representations in the auxiliary space.
The parameter $v$ can be viewed as a ``fugacity'' weighing different symmetric
tensor representations. Finally, from the 2D point of
view the constants $v_n$ manifest themselves as ``horizontal fields''.

The transfer-matrices \eqref{TM} commute, provided they have the same
values of $\mu_1,\mu_2,\mu_3$, \ $\{v_1,\ldots,v_N\}$ \ and
$\{w_1,\ldots,w_N\}$. The proof, of course, follows from
\eqref{YBE-c}, but requires some additional explanations.
There are two places where \eqref{TM} differs from the standard
expression of the transfer matrix. First,
the spectral parameter $w$ is multiplied by different powers of the variable
$q$, depending on the certain sums of $i$- and $j$-type spins. However, as
explained before,
these sums are conserved quantities. Therefore, this modification
merely affects the relative value of the spectral parameter in
different diagonal blocks of the transfer matrix and, obviously,
cannot affect the commutativity.
Further, the factor in the parenthesis in \eqref{TM}, involving the
product over $n$, essentially reduces to the transformation
\eqref{adfield}, since ${\mathcal I}_n^{(M)}$ is a conserved quantity
for the periodic boundary condition in the left-to-right direction.

\subsection{Rank-size duality}

As mentioned above the composite weight \eqref{decomp} decomposes
into an infinite direct sum of the $R$-matrices, corresponding to
symmetric tensor representations of the quantized affine
algebra $U_q(\widehat{sl}_N)$. Therefore the transfer matrix
\eqref{TM} is also an infinite sum of transfer matrices, related to
$U_q(\widehat{sl}_N)$. First, there is a sum over all symmetric tensor
representations $\pi_J$, \ $J=0,1,2,\ldots, \infty$, corresponding to
the auxiliary (horizontal) space and then a direct sum over all
possible representations
\beq
{\mathcal H}=\pi_{{\mathcal I}^{(N)}_1}\otimes\pi_{{\mathcal I}^{(N)}_2}\otimes\cdots
\otimes\pi_{{\mathcal I}^{(N)}_M}
\eeq
in the quantum (vertical) space of the chain
of the length $M$. These representations will be labelled by the
sequence of integers
\beq
I^{(N)}=\{{\mathcal I}^{(N)}_1,{\mathcal I}^{(N)}_2,\ldots,
{\mathcal I}^{(N)}_M\},\qquad {\mathcal I}^{(N)}_m=0,1,2,\ldots
\eeq
where ${\mathcal I}^{(N)}_m$ is defined in \eqref{sum-def}. Then the
transfer matrix \eqref{TM} can be written as\footnote{For simplicity
  we assume here $\mu_1=\mu_2=\mu_3=\mu$.},
\begin{equation}\label{t1}
\boldsymbol{T}(v,w\,|\,\{v\},\{w\}) \;=\; \mathop{\bigoplus}_{\{I^{(N)}\}}\
\sum_{J=0}^\infty\  v^J  \ {\mathbf T}^{sl_N}_{I^{(N)},J}(w\,|\,\{v\},\{w\}),
\end{equation}
where $w$ is the spectral parameter, $\{w\}$ defines its
inhomogeneities, $v$ stands for the ``horizontal fugacity''
and $\{v\}$ defines the horizontal fields.
However, using the symmetry relations \eqref{sym1} and \eqref{sym2}
one can swap the left-to-right and front-to-back directions (and
also reverse the vertical direction) and then rewrite the transfer
matrix \eqref{t1} in the form
\begin{equation}\label{t2}
\boldsymbol{T}(v,w\,|\,\{v\},\{w\}) \;=\; \mathop{\bigoplus}_{\{I^{(M)}\}}\
\sum_{K=0}^\infty\  w^K  \  \Big[{\mathbf T}^{sl_M}_{I^{(M)},K}(v\,|\,\{w\},\{v\})\Big]^T
\end{equation}
where the rank of the algebra $N$ is exchanged with length of the
chain (and vice versa), the spectral parameter is exchanged with the
horizontal fugacity and the set of the spectral parameter
inhomogeneities is exchanged with the set of horizontal field. The
superscript $T$ denotes the transposition in the quantum space and
\beq
I^{(M)}=\{{\mathcal I}^{(M)}_1,{\mathcal I}^{(M)}_2,\ldots,
{\mathcal I}^{(M)}_N\},\qquad {\mathcal I}^{(M)}_n=0,1,2,\ldots
\eeq
where ${\mathcal I}^{(M)}_n$ is defined in \eqref{sum-def}.
This remarkable relation is called the {\em rank-size duality}.
Other instances of this duality were previously found in
\cite{Bazhanov:1992jqa,Bazhanov:2005as}. Somewhat similar phenomena
arise in quantum spin tubes and spin ladders \cite{Batchelor:1999}.
It would be extremely interesting
to understand this duality further, in particular,
to study its implications to the algebraic and analytic
structure of the Bethe Ansatz.

\section{Conclusion}
In this paper we constructed a solution of the tetrahedron equation
which only contains non-negative matrix elements. It is given
explicitly by \eqref{rp7} and \eqref{R-new}. Various properties
of this solution, including symmetry relations, are discussed in
Sect.3. Further, in Sect.4
we have defined a
solvable model of statistical mechanics on a regular cubic lattice
with periodic boundary conditions. Its partition function is given by
\eqref{ZI-def}. The layer-to-layer transfer matrices of the model
form a two-parameter commutative family and possess a remarkable
rank-size duality Eqs.\eqref{t1},\eqref{t2}, previously discovered in
\cite{Bazhanov:1992jqa,Bazhanov:2005as}.

Further properties of the proposed model will be considered elsewhere.
It appears that the constructions of this work give new insights into
the algebraic structure of the 2D integrable models associated with
quantum affine algebra $U_q(\widehat{sl}(N))$. In particular, even in
the simplest case of $N=2$, related to the six-vertex model, one can
obtain many new and rather explicit expressions for associated
solutions of the Yang-Baxter equation. These questions are considered
in our forthcoming paper \cite{BMS13}.

\section*{Acknowledgments}
The authors thank R.J.~Baxter for his interest to the work and
stimulating discussions. This work is partially supported by the
Australian Research Council.

\bibliography{total32m}

\bibliographystyle{utphys}

\end{document}